\documentclass[conference]{IEEEtran}
\IEEEoverridecommandlockouts
\usepackage{cite}
\usepackage{amsmath,amssymb,amsfonts}
\usepackage{algorithmic}
\usepackage{graphicx}
\usepackage{textcomp}
\usepackage{xcolor}
\usepackage{changes}   
\usepackage{svg}

\def\BibTeX{{\rm B\kern-.05em{\sc i\kern-.025em b}\kern-.08em
    T\kern-.1667em\lower.7ex\hbox{E}\kern-.125emX}}
\begin{document}

\title{CSI-Assisted Edge SLAM Testbed Platform for\\ 5G Connected Unmanned Autonomous Vehicles\\

\thanks{This research has been supported by the Ministry of Science, Technological Development and Innovation (Contract No. 451-03-34/2026-03/200156) and the Faculty of Technical Sciences, University of Novi Sad through project “Scientific and Artistic Research Work of Researchers in Teaching and Associate Positions at the Faculty of Technical Sciences, University of Novi Sad 2026” (No. 01-3609/1).}
}

\author{\IEEEauthorblockN{Boris Radovanovic\IEEEauthorrefmark{1}, Sasa Talosi\IEEEauthorrefmark{1}, Srdjan Sobot\IEEEauthorrefmark{1}, Dejan Vukobratovic\IEEEauthorrefmark{1} 
\vspace{1mm}
\IEEEauthorblockA{
\IEEEauthorblockA{\IEEEauthorrefmark{1}Faculty of Technical Sciences, University of Novi Sad, Serbia}
}}}

\maketitle

\begin{abstract}
The evolution from 5G towards 6G reinforces interest in connected robotics, where mobile robots offload compute-intensive tasks to edge servers over ultra-reliable low-latency communication (URLLC) links. Simultaneous localization and mapping (SLAM), a fundamental yet demanding robotics function, is increasingly considered for edge deployment within mobile edge computing (MEC) frameworks. In parallel, integrated sensing and communications (ISAC) enables the use of radio channel information, such as channel state information (CSI), as an additional sensing modality in radio-based SLAM.
In this paper, we design and implement a CSI-assisted Edge SLAM testbed integrating a custom unmanned ground vehicle (UGV), a ROS2-based SLAM framework, and a 5G Open Radio Access Network (O-RAN) system. The proposed architecture provides an end-to-end, cross-layer view of ROS2 sensor data streaming over 5G, explicitly enabling CSI exposure and integration into the SLAM pipeline. We analyze ROS2–DDS communication, RTPS packetization, and 5G user-plane transport, and discuss mechanisms for CSI extraction and delivery via O-RAN components.
The platform enables realistic experimentation with communication-aware SLAM and reveals key challenges related to latency, data streaming, synchronization, and cross-system integration, providing insights for future 6G-enabled robotic platforms.
\end{abstract}


\section{Introduction}

Connected robotics is one of the primary use cases for ultra-reliable and low-latency communication (URLLC) in 5G networks \cite{urllc}, and it continues to be a key driver in the evolution towards hyper-reliable low-latency communications (HRLLC) in 6G \cite{hrllc}. Recent studies have explored use case requirements, system architectures, and standardization efforts for 6G-enabled robotic systems \cite{6Grob1, 6Grob2, 6Gstd1, 6Gstd2}. While early work has often focused on tightly controlled environments or teleoperation-centric applications such as remote surgery \cite{6Gts1}, increasing attention is being directed towards mobile autonomous robots operating in dynamic and unstructured environments. In such settings, robots must continuously perceive, localize, and map their surroundings while meeting stringent latency and reliability constraints, motivating the offloading of compute-intensive perception and decision-making tasks to edge infrastructure over 5G/6G networks \cite{6Grob1, 6Grob2}.

Simultaneous localization and mapping (SLAM) represents a core building block of autonomous mobile robotics, enabling robots to estimate their pose while constructing a map of the environment from onboard sensor data. Modern SLAM pipelines, particularly those relying on high-rate visual or multi-modal sensing, impose significant computational and communication demands that can exceed the capabilities of resource-constrained robotic platforms. This has motivated the concept of Edge SLAM, where sensor data acquired on the robot is streamed over a 5G network to edge servers executing SLAM algorithms within a mobile edge computing (MEC) framework. While this paradigm offers the potential for enhanced performance and reduced onboard complexity, it introduces new challenges related to real-time data streaming, latency constraints, synchronization, and cross-layer interaction between robotic middleware and communication systems.

At the same time, emerging integrated sensing and communications (ISAC) approaches suggest that radio channel measurements, such as channel state information (CSI), can be leveraged as an additional sensing modality for localization and mapping. However, existing studies on edge-based SLAM and radio-assisted localization are predominantly limited to simulation or abstract system models, with limited insight into practical system integration across robotic software frameworks (e.g., ROS2), wireless transport layers, and disaggregated 5G architectures. In particular, the end-to-end realization of CSI-assisted SLAM over a 5G Open RAN (O-RAN) system, including the exposure of radio measurements to robotic applications, remains largely unexplored.

In this paper, we present a comprehensive testbed platform for CSI-assisted Edge SLAM over a 5G O-RAN system, enabling end-to-end experimentation with connected autonomous mobile robots. First, we design and implement a fully integrated system comprising a custom unmanned ground vehicle (UGV) equipped with visual and odometry sensors, a ROS2-based communication framework, and an O-RAN testbed supporting edge deployment of SLAM algorithms. Second, we provide a detailed cross-system and cross-layer analysis of ROS2 data streaming over 5G, with particular focus on DDS-based message serialization, RTPS packetization, and their impact on latency and reliability. We further investigate how O-RAN control mechanisms, including near-real-time RIC, can be leveraged for communication-aware optimization of robotic data flows. Third, we extend the Edge SLAM architecture by introducing CSI as an additional sensing modality, inspired by radio-based SLAM approaches. We explore practical methods for exposing CSI to the SLAM pipeline, including direct extraction at the O-DU and forwarding via a ZMQ–ROS2 bridge, as well as RIC-assisted processing and controlled dissemination through xApps. The proposed platform provides a realistic and flexible environment for studying the interplay between communication and perception in future 5G/6G-enabled robotic systems.

\section{Related Work: 5G-Enabled Edge SLAM}

Simultaneous localization and mapping (SLAM) is a fundamental problem in mobile robotics, concerned with the joint estimation of a robot’s pose and the map of an unknown environment based on sensor observations. Classical and modern SLAM approaches span a wide range of methods, including filter-based, optimization-based, and graph-based techniques, and rely on diverse sensing modalities such as vision, inertial measurements, and odometry. Comprehensive overviews of SLAM algorithms and systems can be found in the literature, forming the basis for recent developments in distributed and communication-aware SLAM \cite{PRbook,slam}.

\textbf{Edge-Assisted SLAM:} With the increasing computational complexity of modern SLAM pipelines, recent works have explored offloading parts of the SLAM process from resource-constrained robots to edge infrastructure. In this context, Edge SLAM has been proposed as a paradigm where computationally intensive components, such as feature extraction, map optimization, or loop closure, are executed at edge servers, while latency-critical tasks remain onboard the robot. For example, Ben Ali et al. propose Edge-SLAM, where visual SLAM is partitioned between the robot and the edge, focusing on adaptive task offloading strategies based on system load and network conditions \cite{BenAli2020}. Similarly, Xu et al. investigate edge-assisted semantic visual SLAM, demonstrating performance gains by offloading semantic processing to edge nodes \cite{Xu2020}. While these works provide important insights into algorithm partitioning and offloading strategies, they largely abstract the underlying communication system and do not explicitly address the impact of wireless transport, protocol behavior, or cross-layer interactions on SLAM performance.

\textbf{Radio-Assisted and ISAC-Enabled SLAM:} In parallel, the integration of communication signals into the SLAM pipeline has emerged as a promising research direction. Recent advances in integrated sensing and communications (ISAC) enable the use of radio channel measurements, such as time-of-arrival, angle-of-arrival, and channel state information (CSI), as additional sensing modalities. This has led to the development of radio-based SLAM approaches, where environmental mapping and localization are enhanced using radio signals. For instance, Kabiri et al. study the fusion of 5G time-of-arrival measurements with inertial data for indoor pose estimation, comparing graph-based and Kalman filter-based approaches \cite{Kabiri2024}. More broadly, recent works have explored the role of CSI and other radio features in localization and mapping, highlighting their potential for improving robustness in challenging environments \cite{Ge2025}. However, these studies are typically limited to algorithmic evaluations or simulations and do not address practical system integration aspects, particularly in the context of real 5G deployments.

\textbf{5G-Enabled Edge SLAM:} Several recent works have begun to investigate the deployment of SLAM over 5G networks. Experimental studies, such as the Ph.D. work in \cite{Sosalla2025}, demonstrate the feasibility of offloading SLAM components over private 5G networks in industrial environments, highlighting benefits in terms of computational scalability and system flexibility. Similarly, other works consider outdoor deployments over commercial 5G macro-cellular networks, evaluating performance under realistic mobility and channel conditions \cite{Karfakis2023}. Despite these advances, existing implementations typically treat the communication system as a black box and do not analyze the interaction between robotic middleware (e.g., ROS2), data serialization mechanisms, and 5G network. In particular, the impact of middleware protocols such as DDS and RTPS on latency, reliability, and bandwidth utilization in 5G-enabled SLAM systems remains largely unexplored.

\textbf{ROS2 Communication over 5G and O-RAN Integration:}
The ROS2 has become the de facto standard middleware for distributed robotic systems, leveraging the Data Distribution Service (DDS) for scalable and flexible communication. Its architecture and communication mechanisms are well documented in the literature, including recent tutorial works \cite{Macenski2022}. Practical integration of ROS2 with 5G networks has been explored in recent industrial collaborations, such as the joint effort between Ericsson and eProsima, which demonstrates the use of DDS-based communication over 5G infrastructures \cite{EP2022}. At the same time, emerging Open RAN (O-RAN) architectures provide new opportunities for fine-grained control and programmability of the radio access network \cite{orantut}. Recent studies have explored mechanisms for exposing radio channel information, including CSI, through O-RAN interfaces and frameworks, enabling new cross-layer applications \cite{Bouknana2025, Polese2026}. However, the integration of such capabilities with robotic middleware and SLAM pipelines remains an open challenge.

\begin{figure*}[!t]
\centering
\includegraphics[width=7in]{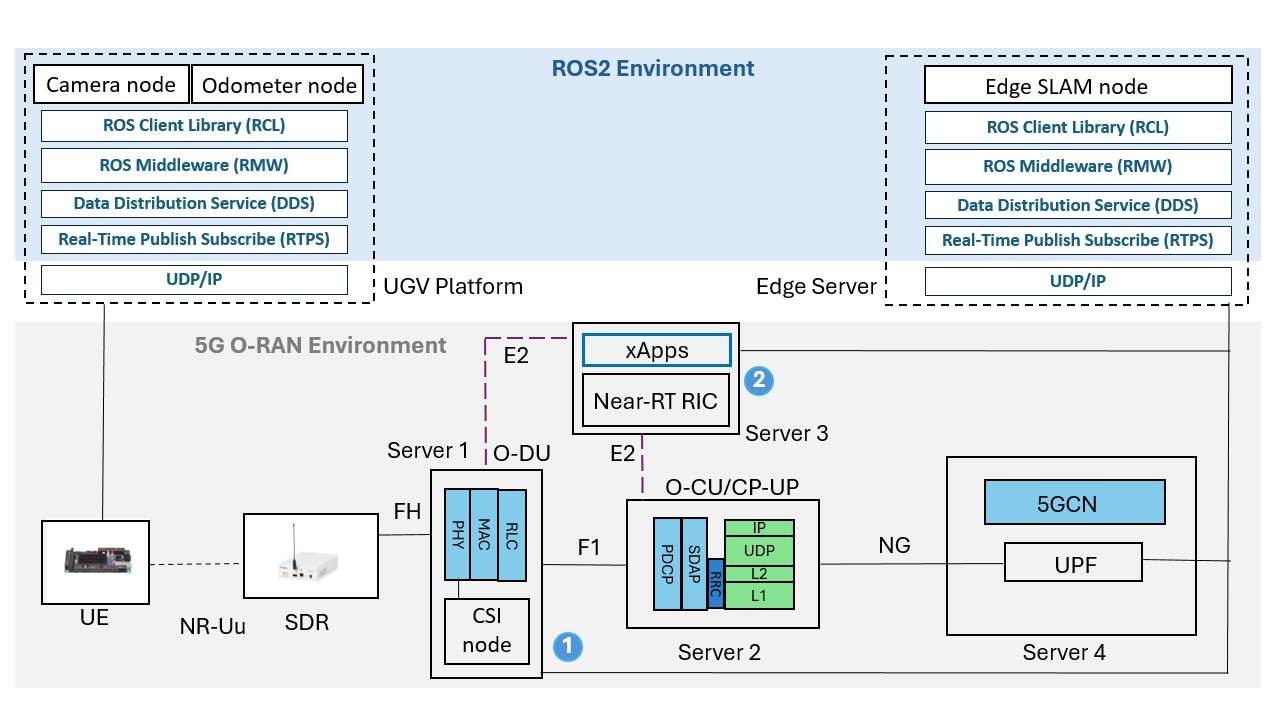}
\caption{CSI-Assisted Edge SLAM over 5G Architecture.}
\label{Fig_1}
\end{figure*}

\section{CSI-Assisted 5G Edge SLAM Architecture}

In contrast to existing studies, this paper provides a comprehensive end-to-end implementation and analysis of a CSI-assisted Edge SLAM system over a 5G O-RAN testbed. Specifically, we i) design and deploy a fully integrated platform combining a ROS2-based robotic system with an in-house 5G O-RAN infrastructure, ii) perform a detailed cross-layer investigation of ROS2–DDS communication over the 5G user plane, and iii) consider practical mechanisms for exposing and integrating CSI into the SLAM pipeline via both direct O-DU extraction and RIC-assisted processing. This holistic approach enables a realistic evaluation of communication–perception interplay and addresses key gaps in current literature regarding system integration, cross-layer behavior, and radio-assisted SLAM in operational 5G environments.

\subsection{Edge SLAM Architecture} \label{Edge SLAM Arch}

The Edge SLAM architecture is based on the ROS2 framework, enabling distributed communication between sensing modules deployed on the UGV and a SLAM node executed at the edge server, as illustrated in Fig. \ref{Fig_1}. The system follows a publish–subscribe paradigm, where onboard sensor nodes generate data streams that are transmitted over the 5G network to the edge-hosted SLAM process.

On the UGV platform, as an example, two primary sensing nodes are considered: a camera node and an odometer node. The camera node periodically acquires image frames and encapsulates them into \emph{sensor\_msgs/Image} messages, while the odometer node produces motion estimates (e.g., incremental pose or velocity) using \emph{nav\_msgs/Odometry} messages. Each node operates as an independent ROS2 publisher, generating time-stamped data streams associated with specific topics.

Messages generated at the application level are processed through the ROS2 communication stack. First, the ROS client library (RCL) provides the API for node creation, topic management, and message handling. The data is then passed to the ROS middleware (RMW) layer, which abstracts the underlying communication implementation. The RMW layer interfaces with a DDS implementation (Fast DDS), which is responsible for data-centric publish–subscribe communication. At the DDS layer, ROS2 messages are serialized into a binary format using the Common Data Representation (CDR). The serialized data is then encapsulated into Real-Time Publish–Subscribe (RTPS) messages, which define the wire protocol used for DDS communication. For large data payloads, such as camera images, RTPS supports fragmentation into multiple submessages, which are subsequently reassembled at the receiver. These RTPS messages are finally transported over standard UDP/IP sockets, forming the interface between the ROS2 environment and the underlying 5G network infrastructure.

\begin{figure*}[!t]
\centering
\includegraphics[width=7in]{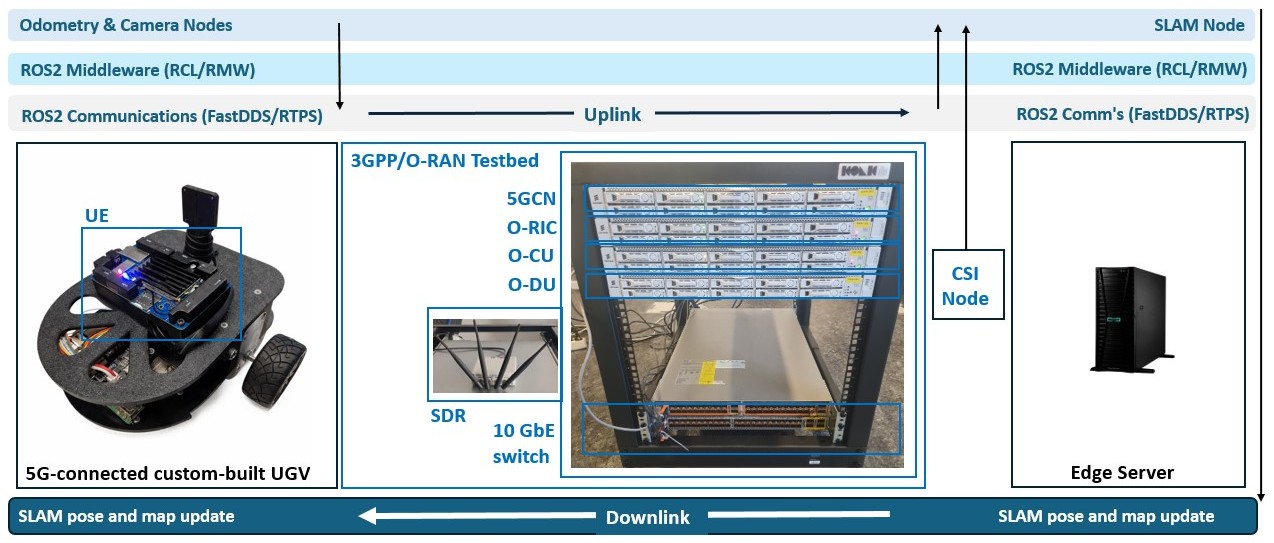}
\caption{Edge SLAM 5G O-RAN Testbed.}
\label{Fig_2}
\end{figure*}

On the edge server, the SLAM node subscribes to the corresponding camera and odometry topics. Incoming UDP/IP packets are processed in the reverse order: RTPS messages are received and reassembled, deserialized into ROS2 message types, and delivered through the RMW and RCL layers to the SLAM node. The SLAM process then performs state estimation and map construction based on the received multi-modal sensor data. The described architecture naturally supports the integration of additional sensing modalities. In particular, a CSI node can be introduced as an additional ROS2 publisher, providing radio channel measurements as a new data stream to the SLAM node. While the mechanisms for CSI extraction and transport are discussed later in this section, from the ROS2 perspective, CSI data can be treated analogously to other sensor inputs, enabling seamless integration into the existing publish–subscribe framework.

\subsection{3GPP/O-RAN 5G Network Architecture}

The 5G testbed considered in this work follows the 3GPP and O-RAN architectural principles, featuring a disaggregated next-generation base station (gNB), as illustrated in Fig. \ref{Fig_1}. The gNB is split into a distributed unit (O-DU) and a centralized unit (O-CU), interconnected via the F1 interface. The O-CU is further divided into control-plane (O-CU-CP) and user-plane (O-CU-UP) functions, enabling flexible deployment and scaling of control and data processing functionalities. On the radio side, the O-DU interfaces with software-defined radio (SDR) units acting as radio units (RUs) through the fronthaul (FH) interface, following 3GPP functional split option 8. Toward the core network, the O-CU connects to the 5G Core Network (5GCN) via the NG interface, comprising NG-C and NG-U for control-plane and user-plane traffic, respectively. The 5GCN follows a service-based architecture, with the user-plane function (UPF) handling user data routing and enabling local breakout towards edge-hosted applications.

In accordance with the O-RAN framework, the radio access network is controlled by a RAN Intelligent Controller (RIC). In this work, we consider the near-real-time (Near-RT) RIC, which interfaces with both O-DU and O-CU via the E2 interface. The Near-RT RIC hosts programmable applications (xApps) that enable dynamic and use-case-specific optimization of radio resource management, including uplink and downlink scheduling policies relevant for latency-sensitive robotic data streams (e.g., odometry and camera data).

From a data transport perspective, ROS2-generated UDP/IP traffic originating from the UGV traverses the 5G user plane, being encapsulated within the 3GPP protocol stack and transported over the NR-Uu interface. After passing through the O-DU, O-CU-UP, and UPF, the traffic is forwarded to the edge server hosting the SLAM node. This end-to-end integration enables seamless coupling between the ROS2-based robotic system and the 5G O-RAN infrastructure, forming the basis for cross-layer optimization and performance analysis.

\subsection{CSI-exposure and integration into SLAM:} Our testbed design goal is to enable the third sensing input to the SLAM node: the CSI data. They can be captured in raw form at the O-DU, however, no 3GPP or O-RAN standard-based option exists to expose these data. Here, we propose two possible directions illustrated as Option 1 and 2 in Fig. \ref{Fig_1}. 

In Option 1, which we have implemented and tested, the CSI data is extracted from O-DU processing pipeline on a subframe basis and stored in a database at O-DU (Server 1). The publish/subscribe messaging mechanism (using ZMQ or similar messaging library) is implemented, along with the ROS2 node wrapper (bridge), to expose the latest database entries, along with the timestamp data, and UE identity, as a new ROS2 CSI node to the subscriber at the edge server. 

In Option 2, which is under development, we are developing an xApp as part of the FlexRIC (Server 3) that would collect CSI data via E2 interface (E2 Service Model) and expose it in a similar way as above. The xApp design could include additional metrics to be exposed (e.g., SNR, RSSI) as well as additional added-value processing on the CSI data. We note that this paper reports only the architecture, but not the actual usage of CSI data in SLAM algorithms, which is the part of our ongoing work.

\section{5G Edge SLAM Testbed}

In this section, we describe an integrated end-to-end Edge SLAM testbed that connects a custom-designed UGV via the 5G O-RAN testbed to edge compute infrastructure. 

\subsection{3GPP/O-RAN 5G Network Testbed Deployment} 
To ensure reproducible end-to-end experiments, we built a compact, self-contained rack that integrates radio, switching, and compute resources. Traffic across the fronthaul, backhaul, and edge planes is kept logically isolated through a single top-of-rack switch shared by four servers.

\subsubsection{Compute and Switching Infrastructure}
Each of the four Quanta D51B-1U servers carries dual Intel Xeon E5-2600 v3 CPUs, 32 GB of RAM, and a 500 GB disk. Server-to-switch uplinks run over 10G SFP+ DAC Twinax cables terminating at a Cisco N5K-C56128P. A dedicated management laptop 
reaches every node over SSH, centralizing control and observation throughout experiments.

\subsubsection{Radio Front-End and User Equipment}
The radio front-end consists of a National Instruments USRP B210 (four antennas) attached to Server 1 over USB 3.0. Quectel RMU500-EK development kit, built around the RM500Q-GL modem, act as 5G endpoint integrated at the UGV (Fig. \ref{Fig_2}), and provides stable, repeatable performance benchmarks.

\subsubsection{Functional Split and Software Stack}
The deployment follows a 3GPP Split 8 topology. Server 1 runs the srsRAN DU~\cite{srsran_project}, configured for TDD operation at 3825~MHz with 2$\times$2 MIMO and 20~MHz bandwidth; Server 2 runs the corresponding srsRAN 
CU~\cite{srsran_project}; Server 4 hosts the Open5GS 5G core network~\cite{open5gs}. FlexRIC is deployed on Server 3, physically and logically decoupled from the fronthaul and backhaul segments to avoid interference during RIC experiments.

\subsection{Edge SLAM Testbed Deployment} 
The Edge SLAM deployment follows the architecture described in Sec. \ref{Edge SLAM Arch}, with ROS2 Jazzy being the middleware on both the UGV and the edge server. The system diagram in Fig. \ref{Fig_1} and Fig. \ref{Fig_2} presents a simplified view of primary data flows. In practice, the ROS2 graph includes additional nodes and topics, such as TF broadcasters, robot state publisher, control manager, etc., which are omitted to emphasize the communication and networking aspects of the system.

\subsubsection{UGV side deployment}
The UGV is equipped with a monocular USB camera, which, in the current setup, serves as a primary source for measurements (i.e. AprilTag \cite{olson2011} detections), as well as DC motors equipped with encoders, which provide odometry. The UGV runs a minimal set of ROS2 nodes responsible for data acquisition and low-level control. As noted in Sec. \ref{Edge SLAM Arch}, a camera node publishes image frames, while the differential drive controller publishes the wheel odometry and the corresponding \emph{"odom to base\_link"} transform via the TF tree. To achieve higher throughput, lower latency, and higher image resolution, the camera node publishes frames as \emph{sensor\_msgs/CompressedImage}. An AprilTag detection node subscribes to this compressed image stream, identifies fiducial markers in the environment, and publishes both detection metadata and the estimated \emph{"base\_link to tag"} transforms. All sensor messages retain their original acquisition timestamps embedded in the message header, which is crucial for correct data association under communication delay.

\subsubsection{Edge server deployment}
The edge server hosts both the AprilTag detection node and the SLAM graph construction node. The detection node subscribes to the compressed image stream transported over the 5G network and publishes detection results on the server. The SLAM graph construction node then subscribes to these local detections as well as to the odometry topic streamed over the 5G network. In Figures  \ref{Fig_1} and \ref{Fig_2}, this is depicted as the Edge SLAM node. The node integrates odometry twist messages to maintain the running pose estimate of the robot while associating landmark observations to the corresponding robot pose using the detection timestamp rather than the message arrival time. Specifically, each landmark observation is matched to the nearest historical pose vertex via timestamp comparison, ensuring that each measurement edge in the graph connects to the pose at which the image was actually captured. This decouples SLAM data association from the network transport latency, making graph construction delay-invariant by design.

\subsubsection{Outlier mitigation}
Monocular AprilTag pose estimation is subject to a well-known pose flip ambiguity inherent to the PnP (Perspective-n-Point) problem, which can produce landmark detections with errors of a few meters despite high detection confidence. This issue is further worsened by network-induced frame rate reduction and JPEG compression artifacts that degrade tag corner localization. To address this, several filtering stages are applied to the graph construction level: (i) rejection of detections during fast robot rotation, where TF timing mismatches are most likely; (ii) a pending landmark buffer that requires multiple agreeing observations from different viewpoints before creating a landmark vertex; (iii) and a median consistency check that rejects any future observations deviating significantly from the established landmark position. At the optimization level, a robust kernel is applied to landmark edges, which limits the influence of any outliers that pass the preceding filters.

\subsubsection{SLAM formulation}
The SLAM problem is formulated as pose graph optimization, where the graph is constructed online during robot operation and optimized offline using iterative Gauss-Newton optimization \cite{PRbook}. We note that offline optimization does not affect the validity of the communication analysis, as the graph construction process, which is the component interacting with the 5G network, is identical in both offline and online formulations. An online variant, in its simplest form, would simply invoke the same optimization periodically during operation.

\section{Preliminary Results}

In this section, we present initial measurements characterizing the 5G transport performance of the testbed and validate the Edge SLAM pipeline through simulation-based experiments with emulated network delay.

\subsection{5G Transport Characterization}
\begin{table}[!t]
\renewcommand{\arraystretch}{1.3}
\caption{USB camera stream end-to-end transport delay}
\label{delay_table}
\centering
\begin{tabular}{c||c|c|c|c}
\hline
\bfseries Image Resolution & \bfseries Avg. delay & \bfseries Min. delay & \bfseries Max. delay & \bfseries Std. dev.\\ 
\hline\hline
640x480 pixels & 45 ms & 13 ms & 160 ms & 22 ms\\
\hline
1280x720 pixels & 44 ms & 19 ms & 281 ms & 20 ms\\
\hline
1920x1080 pixels & 67 ms & 35 ms & 175 ms & 17 ms\\
\hline
\end{tabular}
\end{table}

Table \ref{delay_table} summarizes the end-to-end transport delay over the 5G O-RAN testbed for a compressed USB camera stream. Measurements were obtained by recording the camera stream on the UGV using \emph{rosbag2} and replaying it over the 5G link while measuring the one-way delay on the edge server. This isolates the network transport delay from any camera driver and compression latency. All evaluated resolutions operated on the UGV at a frame rate of 30 FPS, with the exception of the 1920x1080 configuration, where the USB camera was limited to a maximum of 15 FPS. The server-side measurements reflected identical frame rates, indicating that no degradation was induced by the network. These results represents a controlled single-robot, low-traffic baseline.

\subsection{Edge SLAM validation}
To validate the delay-invariant property of the Edge SLAM architecture, we conducted experiments using a Gazebo-based digital twin of the UGV and warehouse environment, with emulated network delay at the application layer. A configurable delay node buffers compressed image messages, introducing a base delay of 1 second with uniform random jitter of $\pm300$ ms before forwarding to the AprilTag detection node. This substantially exceeds the delays measured on the physical testbed in order to stress-test the system under adverse conditions representative of congested or degraded network scenarios.

\begin{figure}[!t]
\centering
\includegraphics[width=3.5in]{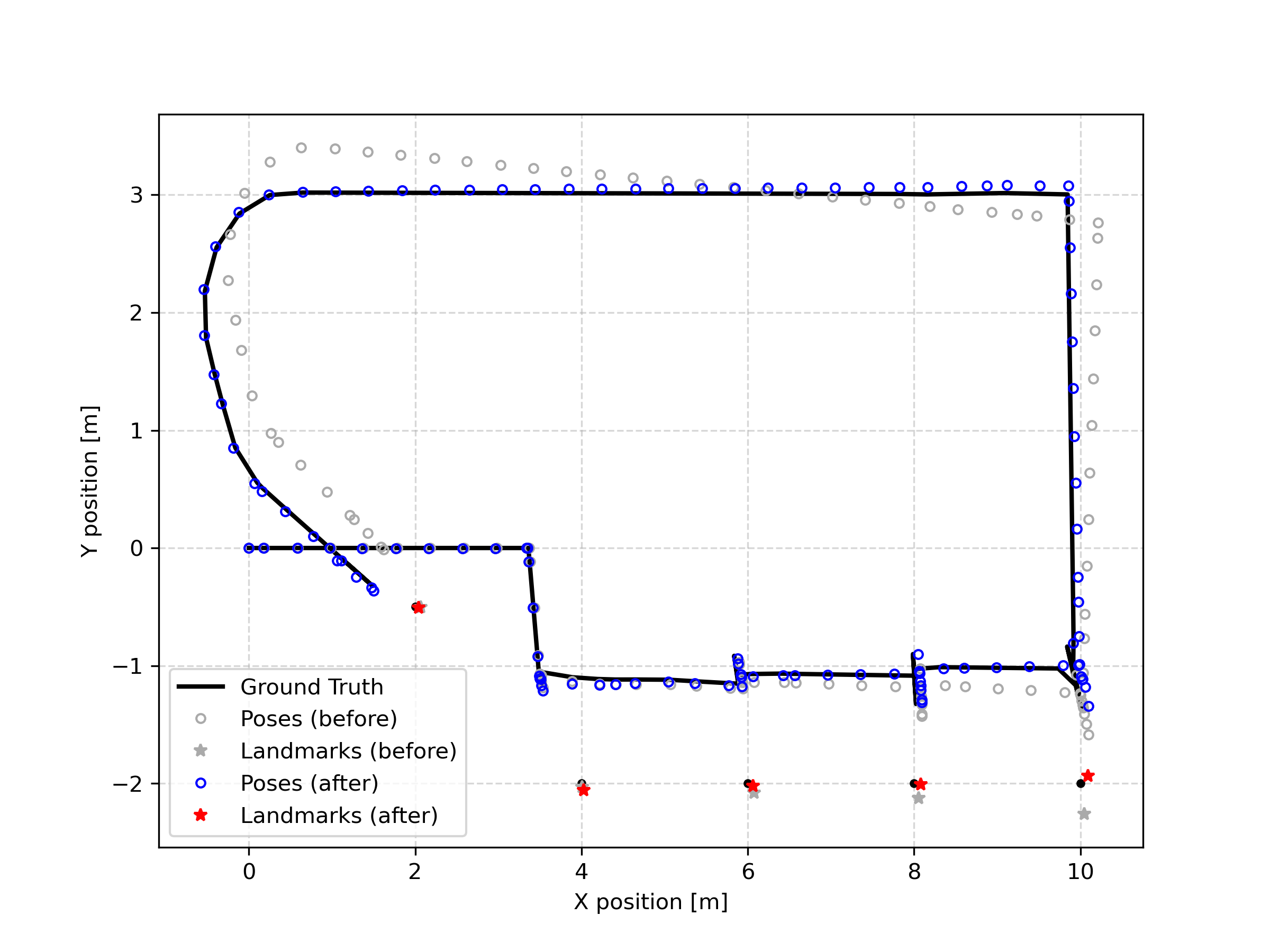}
\caption{Graph SLAM - Before and After Optimization}
\label{Fig_3}
\end{figure}

Fig. \ref{Fig_3} shows the pose graph before and after optimization for a representative run with the emulated delay. Poses (circles) and landmarks (stars) prior to optimization are shown in gray, while optimized estimates are depicted in blue (poses) and red (landmarks). After optimization, the trajectory converges to closely match the ground truth, demonstrating that offloading SLAM to the edge is robust to the 5G network testbed performance and measured delays. Although only a preliminary study, this work serves to stimulate further work and discussion in the domain of 5G/6G connected mobile robotics systems.

\section{Conclusions}

In this paper, we presented our testbed deployment and integration study for Edge SLAM algorithm offloaded from 5G-connected UGV. The focus of the work was not in the domain of SLAM algorithms per se, but in establishing 5G O-RAN testbed suitable for SLAM experimentation. In particular, following the trends of ISAC integration in beyond 5G, we have discussed and (partially) implemented CSI exposure mechanism that could serve as a new ROS2 CSI sensing input to CSI-enhanced SLAM.

\end{document}